
\def\.{\mathaccent 95}

\def\frac#1#2{{\textstyle{{#1}\over {#2}}}}

\def\lsim{\mathrel{\rlap{\lower4pt\hbox{\hskip1pt$\sim$}}
    \raise1pt\hbox{$<$}}}
\def\gsim{\mathrel{\rlap{\lower4pt\hbox{\hskip1pt$\sim$}}
    \raise1pt\hbox{$>$}}}
\def\sqr#1#2{{\vcenter{\vbox{\hrule height.#2pt
         \hbox{\vrule width.#2pt height#1pt \kern#1pt
         \vrule width.#2pt}
         \hrule height.#2pt}}}}

\newbox\grsign \setbox\grsign=\hbox{$>$} \newdimen\grdimen \grdimen=\ht\grsign
\newbox\simlessbox \newbox\simgreatbox
\setbox\simgreatbox=\hbox{\raise.5ex\hbox{$>$}\llap
     {\lower.5ex\hbox{$\sim$}}}\ht1=\grdimen\dp1=0pt
\setbox\simlessbox=\hbox{\raise.5ex\hbox{$<$}\llap
     {\lower.5ex\hbox{$\sim$}}}\ht2=\grdimen\dp2=0pt

%
%

\def\ref#1  {\noindent \hangindent=24.0pt \hangafter=1 {#1} \par}
\def\doublespace {\smallskipamount=6pt plus2pt minus2pt
                  \medskipamount=12pt plus4pt minus4pt
                  \bigskipamount=24pt plus8pt minus8pt
                  \normalbaselineskip=24pt plus0pt minus0pt
                  \normallineskip=2pt
                  \normallineskiplimit=0pt
                  \jot=6pt
                  {\def\smallskip {\vskip\smallskipamount}}
                  {\def\medskip   {\vskip\medskipamount}}
                  {\def\bigskip   {\vskip\bigskipamount}}
                  {\setbox\strutbox=\hbox{\vrule
                    height17.0pt depth7.0pt width 0pt}}
                  \parskip 12.0pt
                  \normalbaselines}

\magnification=1200

$$\bf{OHM'S\ LAW\ FOR\ A\ RELATIVISTIC\ PAIR\ PLASMA}$$
\centerline{Eric G. Blackman and George B. Field}
\centerline{Harvard University}
\centerline{and}
\centerline{Harvard-Smithsonian Center for Astrophysics}
\centerline{60 Garden St., Cambridge, MA, 02138}
\centerline{[published in Phys. Rev. Lett. $\bf 71$, 3481, (1993)].}
\doublespace
\bigskip
$$\bf{ABSTRACT}$$

We derive the fully relativistic Ohm's law for an electron-positron plasma.
The absence of non-resistive terms in Ohm's law and the natural substitution of
the 4-velocity for the velocity flux in the relativistic bulk plasma equations
do not require the field gradient length scale to be much larger than the
lepton inertial lengths, or the existence of a frame in which the distribution
functions are isotropic.

\bigskip

\vfill
\eject
For a plasma, Ohm's law describes the relation between the induced current and
the plasma electric field.  For an ion-electron plasma, the field depends on
resistive, inertial, and Hall effect contributions.  The result is usually
derived for the non-relativistic limit in the Boltzmann picture$^{\rm 1}$.
  Ohm's law plays a direct role in the magnetic induction equation used in the
description of bulk plasma dynamics.

Relativistic plasma models have been effective in explaining the observations
of relativistic astrophysical jets and winds $^{\rm 2}$.
Such models have generally employed the relativistic continuity equation as a
vanishing 4-divergence of the bulk 4-velocity, and Ohm's law as a simple bulk
4-vector generalization of the non-relativistic equation$^{\rm 2,3}$.

We shall see that for a two-component relativistic plasma composed of different
mass particles, the natural use of these magnetohydrodynamic (MHD) forms for
the continuity equation and Ohm's law requires the existence of a reference
frame in which both distribution functions are isotropic in momentum.
The constraint results from the non-linear relation between momenta and
velocities in the relativistic regime.

This requirement is non-trivial because distribution function isotropy also
requires the plasma under study to be microinstability saturated; otherwise
microinstabilities
could grow because of distribution function anisotropy.  Yet, evidence for the
presence of anisotropies and microinstabilities in relativistic winds has come
from observations of the Crab Nebula$^{\rm 4}$.
In particular, the Weibel instability has been suggested to explain the
presence of wisps downstream from relativistic shocks$^{\rm 5}$. Anisotropies
downstream from relativistic shocks are likely present in jets as well. $^{\rm
6}$

Depending on density and temperature conditions, anisotropies in relativistic
plasmas may also arise from anisotropic radiation onto a plasma, such as in
coronae of stars or in models active galactic nuclei (AGN) for which a pair
atmosphere forms.$^{\rm 7}$  The impinging of winds and jets onto ambient media
also produces anisotropies$^{\rm 8}$.

In this paper we show that in contrast to an ion-electron plasma: 1) only a
resistive contribution to Ohm's law for a relativistic $\rm {e^+-e^-}$ plasma
is relevant under quite general conditions, and 2) that anisotropy in the
distribution functions need not affect the form of the unperturbed relativistic
bulk equations for the pair plasma.  Thus, relativistic bulk dynamics for pair
plasmas which exhibit evidence for microinstabilities are appropriately
described by the relativistic MHD formalism, whereas ion-electron plasmas which
exhibit such instabilities are not.


The Boltzmann equation is given by
$$\partial f/ \partial t+v^i\partial_i f+ F^i
\partial_{p_i}f=[\Delta_t f]_{coll.} +[\Delta_t f]_{radiat.}+[\Delta_t
f]_{creation}+[\Delta_t f ]_{annihilation}\eqno(1)$$
where $v^i$ is the particle velocity, $p^i$ is the particle momentum, $t$ is
the time, $F^i$ is the electromagnetic force (since we ignore gravity),
$f=f(\bf x \rm,\bf p \rm,t)$ is the scalar distribution
function, and the terms on the right are schematic.
We shall assume that
collisional losses dominate synchrotron radiation losses, which is acceptable
for $^{10}$
$$d|{\bf p}|/dt|_{synch} / |e {\bf v} {\rm x} {\bf B} {\rm |} << 10^{-16}
\gamma^2 B sin \phi,\eqno(2)$$
where $\gamma$ is the particle Lorentz factor,
$\phi$ is the pitch angle, $e$ is the positron charge, and $B$ is the magnetic
field measured in Gauss.

Define the plasma quantities:
number density
$$n_{\pm} =\int f_{\pm} ({\bf x} , {\bf p},t) d^3p,\eqno(3)$$
 and velocity flux density
$$\phi^i_{\pm} =\int {v^i_{\pm}} f_{\pm} ({\bf x , p}, t)d^3p=n_{\pm}\langle
v^i_{\pm}\rangle.\eqno(4)$$
Adopting the signature $(+,-,-,-)$, we then have the flux density 4-vector $
\phi_{\pm}^{\mu} =(cn_{\pm},\phi^i_{\pm}),$ and current density 4-vector
$j^{\mu} =e({\phi^{\mu}}_+-{\phi^{\mu}}_-).$
The
energy and momentum densities are the $(00)$ and $(i0)$ components of the
symmetric kinetic tensor given by
$$K^{i k}_{\pm}=\int v^i_{\pm}p^k_{\pm}f_{\pm}d^3p=n_{\pm}\langle
v^i_{\pm}p^k_{\pm}\rangle \eqno(5)$$
and
$$K_{\pm}^{0 \mu}=(\epsilon_{\pm}, c\Pi_{\pm}^i),\eqno(6)$$
where $\Pi^i_{\pm}$ is the momentum density of either positrons or electrons.

Quantities without the ${\pm}$ shall refer to the plasma as a whole--
the sum of the component contributions.
We require the existence of a proper frame moving with 4-velocity $U_{\mu}$
in which the charge density, the velocity flux density, and the momentum
density vanish.  We denote quantities in this frame with a superscript *.  For
an $e^+-e^-$ plasma this means
$$j^*_0=e(n_+^*-n_-^*)=0,\eqno(7a)$$
$${\phi^{i*}}=0,\eqno(7b)$$
and
$${\Pi^i}^*=m_e(n^*/2)(\langle \gamma_+v^i_+\rangle ^*+\langle
\gamma_-v^i_-\rangle^*)=0.\eqno(7c)$$
Since $K_{\mu \nu}$ has 10 independent components,
we can write
$$K_{\mu \nu}^*=P_{\mu \nu}^*+A_{\mu \nu}^*,\eqno(8)$$
where the symmetric tensors $P_{\mu \nu}$ and $A_{\mu \nu}$ satisfy
$$P_{ij}^*=P \delta_{ij}\ ,\ P_{0i}=0,\ P_{00}=\epsilon^*,\eqno(9)$$
and
$$A_{0\mu}^*=0. \eqno(10)$$
In $(8)\ -\ (10)$, $P$ is the scalar pressure and $A^*_{ij}$ has 5 independent
components that measure anisotropy.

Finally, define the 4-vector $H^{\mu}$ by
$$H^{\mu}=[2/(m_++m_-)][\rho^*U^{\mu}-(m_+\phi_+^{\mu} +
m_-\phi^{\mu}_-)].\eqno(11)$$
where $\rho^*$ is the proper frame rest mass density.
For a pair plasma this becomes
$$H^{\mu}_{pair}=(1/m_e)][\rho^*U^{\mu}-m_e(\phi^{\mu})].\eqno(12)$$
Thus from $(7b)$,  ${H^i}^*_{pair}$ vanishes.  But ${H^0}^*_{pair}=0$
by definition, so ${H_{pair}^{\mu}}^*=0$.  Since $H^{\mu}_{pair}$ is a
4-vector, the vanishing of ${H^{\mu}}^*_{pair}$
implies that $H^{\mu}_{pair}=0$ in all frames.  Note that ${H^{\mu}}^*$
measures the heat flux density per unit mass in the proper frame, so we shall
call $H^{\mu}$ the heat flux density 4-vector.

The procedure used to derive Ohm's law for a relativistic pair
plasma is as follows:
 (i) First we obtain a ``resistive'' type
collision term appropriate for a nearly collisionless relativistic pair plasma.
(ii) Second, we relate this to the current density.
(iii) Finally, a subtraction of the momentum density equations for the
positrons and electrons yields the desired result.


We can find the value of the momentum density in any frame by Lorentz
transforming the
stress-energy tensor.  The result is
$$c\Pi^i=c^2m_e(n_+\langle u_+^i\rangle +n_-\langle u_-^i\rangle )
=\gamma_V^2(P+\epsilon^*)
V^ic^{-1}+A^{0i},\eqno(13)$$
where $u^i$ is a spatial component of the particle
4-velocity, $V^i$ is a component of the bulk 3-velocity, $\gamma_V$ is the bulk
Lorentz factor, and the anisotropy term is given by
$$A_{0i}=A^*_{ij}U^j+U^kU^lA_{kl}^*U^i/(\gamma_V+1).\eqno(14)$$
Note that since the proper frame 4-vector ${A^{\mu \nu}}^*U_{\nu}^*=0$, we know
that it is zero in all frames.  Thus $A_{i0}=A_{ik}U^k$
and $(13)$ can be written
$$c\Pi^i=c^2m_e(n_+\langle u_+^i\rangle +n_-\langle u_-^i\rangle
)=\gamma_V(P+\epsilon^*)U^i+A^{ij}U_j.\eqno(15)$$
Inverting $(15)$ we obtain
$$\gamma_V\ V^ic^{-1}=U^i=[c^2m_e(n_+\langle u_+^i\rangle +n_-\langle
u_-^i\rangle )-A^{ij}U_j]/[(\gamma_V)(P+\epsilon^*)].\eqno(16)$$

The average 4-momentum gains from collisions in the proper frame are given by
$$\Delta {p_+^{\mu}}^*=-\Delta {p_-^{\mu}}^*=(m_e/2)(\langle u_+^{\mu}\rangle
^*-\langle u_-^{\mu}\rangle ^*).\eqno(17)$$
The approximate proper frame pair plasma collision term is then
$$(n^*\nu_c^*/4)(\Delta{p_+^{\mu}}^*)=-(n^*\nu_c^*/4) (\Delta{p_-^{\mu}}^*)
\equiv{P_{+-}^{\mu}}^*=-{P_{-+}^{\mu}}^*,\eqno(18)$$
where $\nu_c^*$ is the proper frame collision frequency.

Equations  $(12)$ and $(16)$ give
$$m_ec^3(n_+\langle u_+^{i}\rangle +n_-\langle u_-^{i}\rangle
)-cA^{ij}U_j=\gamma_V(P+\epsilon^*){n^*}^{-1}(\phi_-^{i} +
\phi_+^{i}).\eqno(19)$$
Now in the electron frame, we have
$$c^2{\Pi_+^{i}}^{(-)}-[cA^{ij}U_j]^{(-)}=\gamma_V^{(-)}(P+\epsilon^*){n^*}^{-1}{\phi^i_+}^{(-)},\eqno(20)$$
and in the positron frame
$$c^2{\Pi_-^{i}}^{(+)}-[cA^{ij}U_j]^{(+)}=\gamma_V^{(+)}(P+\epsilon^*){n^*}^{-1}{\phi^i_-}^{(+)}.\eqno(21)$$
Transforming $(20)$ and $(21)$ to the proper frame, subtracting, and using
$(7)$ and $(18)$ gives
$$\nu^*_c({\Pi^i}^*_+-{\Pi^i}^*_-)=2{P^i}^*_{\pm}=n^*e\eta_r{j^i}^*,\eqno(22)$$
where the effective resistivity $\eta_r$ is given by
$$\eta_r=\nu_c^*(n^*e)^{-1}c^{-2}\{(P+\epsilon^*)(n^*e)^{-1}[{\gamma^*_c}^2(\gamma^*_c+1)^{-1}j^{\mu *}j_{\mu^*}
(en^*c)^{-2}+\gamma^*_c+1]$$
$$-\gamma^*_c
\epsilon^*(en^*)^{-1}-2{\nu_c^*}^{-1}{\gamma^*_c}^2(\gamma^*_c+1)^{-1}(n^*e)^{-2}{j_{\mu}}^*{P^{\mu}_{\pm}}^*\},\eqno(23)$$
and $\gamma_c^*$ is the Lorentz factor corresponding to the velocity
$$(n^*/2)^{-1}{\phi^i}^*_+=-(n^*/2)^{-1}{\phi^i}^*_-={j^i}^*/(en^*).\eqno(24)$$
We have used the 4-vector indices in $\eta_r$ since ${j_0}^*=0$.


Equation $(22)$ suggests that we subtract the momentum equations for
the electrons
and the positrons to obtain Ohm's law.  This is standard in the
non-relativistic case, but is only fruitful in the relativistic case because
$H^{\mu}_{pair}=0$.

Under the conditions of $(7)$, conservation of energy and momentum
momentum implies that the
contributions of the annilihation and creation terms of $(1)$ to the
the energy and momentum equation cancel for each plasma component in the proper
frame.  Thus, by covariance, the contributions to the 1st moments in all frames
vanish, and we do not consider them further.

The $i$th component of the proper frame relativistic energy mometum tensor for
positrons, as
obtained from the 1st moment of the Boltzmann equation is given by
$$\partial_0{\Pi_+^i}^*=-\partial_k {K_+^{ki}}^*-e(n^*/2){E^i}^* -{[e\langle
{\bf v}_+/c\rangle \ {\rm x}\
{\bf B}]^i}^*+{P_{+-}^i}^*.\eqno(25)$$
Subtracting the analagous equation for electrons, and using $(8)$, $(18)$, and
$(22)$ we get
$$E^{*i}=\eta_rj^{i*}-m_ec(2e\rho^*)^{-1}\partial_k({A_+^{ki}}^*-{A_-^{ki}}^*)-\partial_0[(\eta_r/\nu_c^*) {j^i}^*].\eqno(26)$$

The requirements $(7a)$ and $(7b)$ which led to $(26)$, are assumptions about
the 0th and 1st moments of the Boltzmann equation.  If we further assume that
$${A^{ij}_+}^*={A^{ij}_-}^*,\eqno(27)$$
the second term on the right hand side of $(26)$ would vanish.
Then, in the steady state, we would be left with
$$E^{*i}=\eta_rj^{i*}.\eqno(28)$$

Note that the dependence on the current in $(28)$,
results from $(22)$.  The latter follows for example, even in the Fokker-Planck
approximation if, as in our case, the charge density vanishes in the proper
frame.  In addition, note that none of the assumptions that led to $(28)$
require the existence of a reference frame in which the distribution functions
of positrons and electrons
are isotropic.

In a general frame, $(28)$ becomes
$$F^{\mu \nu}U_{\nu}=\eta_r (j^{\mu}+j^{\tau}U_{\tau}U^{\mu}),\eqno(29)$$
where we have added a convection term.
Equation $(29)$ resembles the magnetofluid
relation $^{11}$.
Here however, the effective resistivity $\eta_r$ depends on a scalar function
of the current and momentum density of the plasma components.
This dependence is eliminated when we choose the coordinate axes such that
the proper frame velocity lies on a principal axis.  In addition,
$\gamma_c^*$ factors out of $(23)$ and we have
$$\eta_r \rightarrow \eta_r^{(pr)}=\nu^*_c
c^{-2}(n^*e)^{-2}\{2(P+\epsilon^*)-\epsilon^*\}=\nu^*_c
c^{-2}(n^*e)^{-2}(2P+\epsilon^*),\eqno(30)$$
where the superscript $(pr)$ indicates that the current flows along a principal
axis.

Note that unlike the Ohm's law for an ion-electron plasma, there are no Hall
effect or pressure contributions to $(29)$.
The vanishing of the bulk Hall effect term is the result of an effectively zero
net particle gyration frequency; the sum of the electron and positron
contributions vanish due to opposite streaming of positrons and electrons
around the field lines.  The vanishing of the pressure term results because our
assumptions have appealed to the mass symmetry of the problem, eliminating
diffusion.
The Hall effect and pressure terms are negligible
for an ion-electron plasma only when $\lambda_{fg}>> \lambda_{i},$ and
$\lambda_{fg}>>\lambda_{i}\beta,$ respectively, where $\lambda_{fg}$ is the
length scale of the field gradients, $\lambda_i$ is the ion-inertial length,
and $\beta$ is the ratio of thermal to magnetic pressure.

The vanishing of the heat flux, which led to $(22)$, is also the condition
which allows substitution of the bulk 4-velocity components for the bulk 4-flux
components in the plasma continuity equation.  The point is that the usual
integration of $(1)$ over momentum gives
$\partial_{\mu}\phi^{\mu}=0$, so substitution of the 4-velocity for the 4-flux
$\phi$ requires $H^{\mu}=0$.
 If $ H^{\mu}\ne 0$, then this substitution would produce an inhomogeneous
equation.  Because $m_+>>m_-$, for a relativistic ion-electron plasma,
${H^{\mu}}^*$ only naturally vanishes if both distribution functions are
isotropic in the proper frame.  This is true
whether or not the proton component has a relativistic temperature.

Note that $H^{\mu}$
vanishes for $any$ two component plasma in the non-relativistic limit.  To see
this note
that
$${H^i}^* \propto -m_+n_+^*\langle v_+^i\rangle ^*+-m_-n_-^*\langle
v_-^i\rangle ^*=m_+n_+^*\langle \gamma_+v_+^i-v_+^i\rangle
^*+m_-n_-^*\langle\gamma_-v_-^i-v_-^i\rangle ^*,\eqno(31)$$
where the second equality follows since the proper frame momentum density
vanishes by definition.
  Thus when $n_+^*=n_-^*$, $(31)$ vanishes when $\gamma_+=\gamma_- =1$.
Therefore $H^{\mu}$ vanishes by the argument that follows equation $(12)$.

Finally, note that our proper frame result is not the full nonrelativistic
limit, because of the relativistic temperature.
In the non-relativistic limit, we have the additional result that the pressure
drops out of $(23)$ and that $\epsilon^*=n^*m_ec^2$ so
$$\eta_r^{(pr)} \rightarrow \eta=(\nu_c n^* m_e c^2)/(n^* ec)^2\eqno(32)$$
$$=m_e\nu_c/(n^*e^2)=2\mu_{pair}\nu_c/(n^*e^2),\eqno(33)$$
where $\eta$ is the non-relativistic resistivity, and $\mu_{pair}\equiv m_e/2$
is the reduced mass for the pair plasma.  Equation $(33)$
 is the usual non-relativistic result.  In the case of a non-relativistic
ion-electron
plasma, $\mu_{pair}$ is replaced by $\mu_{ie}\equiv m_im_e/(m_i+m_e)\sim m_e$.

We would like to thank A. Loeb for discussion.
$$---------------$$
\noindent [1] D. B. Melrose, $Instabilities\ in\ Space\ and\ Laboratory\
Plasmas$,
(Cambridge, Cambridge UK, 1986).

\noindent [2] M. Hoshino et al., Ap. J., $\bf 390$, 454, (1992); R. V. E.
Lovelace et al., Ap. J., {\bf 315}, 504, (1987); S. Appl and M. Camenzind, A.
and Ap., {\bf 206}, 258, (1988); F. C. Michel, Rev. Mod. Phys., {\bf 54}, 1,
(1982).

\noindent [3]
A. Lichnerowicz, $Relativistic\ Magnetohydrodynamics$, (Benjamin, New York,
1967).

\noindent [4]
F. C. Michel et al., Ap. J., {\bf 368}, 463, (1990).

\noindent [5] Y. A. Gallant et al., University of California Institute of
Geophysics and Planetary Physics, preprint UCRL-JC-111254, 1992.

\noindent [6]
J. G. Kirk, in $Extragalactic\ Radio\ Sources--From\ Beams\ to\ Jets$
edited by J. Roland et al., (Cambridge, Cambridge UK, 1992).


\noindent [7] A. P. Lightman and A. A. Zdziarski, Ap. J., {\bf 319}, 643,
(1987).

\noindent [8]
Ya. I. Istomin, in, $Extragalactic\ Radio\ Sources--From\ Beams\ to\ Jets$,
edited by J. Roland et al., (Cambridge, Cambridge UK, 1992).

\noindent [9] V. B. Berestetskii, E.M. Lifshitz, and L. P. Pitaevskii,
$Quantum\ Electrodynamics$, (Wiley, New York, 1979).

\noindent [10]
G. Rybicki and A. Lightman, $Radiative\ Processes$, (Wiley, New York, 1979).

\noindent [11]
J. D. Jackson, $Classical\ Electrodynamics$, (Wiley, New York, 1975).

\end